\documentstyle[prl,aps]{revtex}

\input epsf
\begin{document}

\tighten

\def\be{\begin{equation}}
\def\ee{\end{equation}}

\draft

\preprint{\vbox{
\hbox{CWRU-P7-1998}
}}

\title{$Z_3$ Strings and their Interactions}

\author{Junseong Heo$^{\dag}$$^{\ddag}$ and Tanmay Vachaspati$^{\ddag}$}

\address{
$^{\dag}$ Physics Department, Yale University, New Haven, CT 06511.\\
$^{\ddag}$ Physics Department, Case Western Reserve University,
Cleveland, OH 44106-7079.}

\twocolumn[
\maketitle			

\begin{abstract}
\widetext

We construct $Z_3$ vortex solutions in a model in which
$SU(3)$ is spontaneously broken to $Z_3$. The model is 
truncated to one in which there are only two dimensionless 
free parameters and the interaction of vortices within
this restricted set of models is studied numerically. 
We find that there is a curve in the two dimensional space 
of parameters for which the energy of two asymptotically separated 
vortices equals the energy of the vortices at vanishing separation.
This suggests that the inter-vortex potential for $Z_3$ strings 
might be flat for these couplings, much like the case of $U(1)$ strings
in the Bogomolnyi limit. However, we argue that the intervortex
potential is attractive at short distances and repulsive at large
separations leading to the possibility of unstable bound states of 
$Z_3$ vortices.

\

\end{abstract}


]

\narrowtext

\smallskip

\section{Introduction}
\label{intro}

Vortex solutions are well-studied in a wide variety of condensed
matter systems. In superconductors, vortex solutions have been
recognized since Abrikosov's seminal work \cite{abrikosov}, while
string solutions in relativistic field theory were first found by
Nielsen and Olesen \cite{hnpo}. The interaction of vortices has
also been a subject of continual investigation, starting from the
work of Abrikisov who conjectured a vortex lattice due to the
repulsive force between vortices. In relativistic models, early
work on the interaction of vortices was carried out by Jacobs and
Rebbi \cite{ljcr} in which they noted a transition from attractive
to repulsive interaction as a certain parameter
was varied. Furthermore, at a critical point in parameter space, 
the inter-vortex potential was found to be flat and this is the so 
called Bogomolnyi limit \cite{bogo}.

The investigations thus far have mostly considered the interaction of 
$U(1)$ vortices - the kind that commonly occur in superconductors. However,
there is a much wider variety of vortices occurring in other condensed
matter systems and there is a possibility that these may also exist
in particle physics and cosmology. In particular there is a class of
vortices called ``$Z_N$ vortices'' in which $N$ vortices are topologically
equivalent to the vacuum. The simplest of these is the $Z_2$ (global)
vortex that exists in nematic liquid crystals. $Z_4$ vortices can
be found in the A-phase of He$^3$. To our
knowledge, $Z_3$ vortices have not yet been observed but it is possible
that these may be relevant to confinement in QCD.
This is apparent in the dual standard model picture that one of us
has proposed \cite{tvdual} and is indicated by ongoing
work on supersymmetric dualities \cite{strassler}.

In this paper we study $Z_3$ vortices and their interaction. The symmetry
breaking pattern we consider is
\be
SU(3) \rightarrow Z_3
\label{symbreak}
\ee
which can be accomplished by the vacuum expectation value (VEV) of three
adjoint scalar fields. These details are provided in Sec. \ref{themodel}.
The vacuum manifold of the model is
$$
\Sigma = {{SU(3)} \over {Z_3}} 
$$
and since,
$$
\pi _1 (\Sigma ) = Z_3 \ ,
$$
the model admits $Z_3$ strings which we explicitly construct in 
Sec. \ref{solutions}. The interaction of these strings is studied in
Sec. \ref{interact} by comparing the energy of infinitely separated
(very distant) strings with the energy of the strings at vanishing
separation. 

An interesting result that we obtain is that there is a surface in
parameter space such that the energy 
of two vortices at infinite and at vanishing
separation are equal. This raises the possibility that perhaps the
inter-vortex potential is flat for these values of the parameters,
much like the Bogomolnyi case for $U(1)$ strings. However, we argue 
that the $Z_3$ inter-vortex potential is not flat but has a maximum 
at some finite vortex separation. This 
then indicates that there must exist an unstable, static, bound state 
of two separated $Z_3$ vortices.

In an early paper, de Vega and Schaposhnik \cite{dvs} investigated the
properties of $Z_3$ strings. The group theoretic formalism developed
there is very general and can be used to construct $Z_N$ strings for
arbitrary $N$. The
study of the properties of $Z_3$ strings was, however, restricted to 
a choice of
parameters where the $Z_3$ strings are effectively identical 
to $U(1)$ strings. Furthermore, for this choice of parameters, the
desired symmetry breaking pattern (\ref{symbreak}) is not 
uniquely picked out by
the potential. That is, the $Z_3$ symmetric vacuum is degenerate with
vacuua having other symmetries and is no longer the unique ground state
of the model. 
We explain these comments in more detail in Sec. \ref{critical}.

\section{Model}
\label{themodel}

The construction of a model exhibiting $SU(3) \rightarrow Z_3$
involves the following two steps.
First, identify the necessary ingredients of the model. In
particular, determine the scalar field content of the model.
Second, construct the most general scalar field potential and
determine the range of parameters which lead to the
desired symmetry breaking.
We treat these steps in the next two subsections.

\subsection{Ingredients}

A scalar field, $\Phi$, in the adjoint representation of $SU(3)$
can be written as the $3\times 3$ Hermitian matrix
$$
\Phi = \sum_{a=1}^8 \Phi^a \lambda_a
$$
where, $\lambda_a$ are the Gell-Mann matrices \cite{georgi} and
$\Phi^a$ are 8 real scalar fields. Under the action of $g \in SU(3)$,
$\Phi$ transforms as:
$$
\Phi \rightarrow \Phi ' = g \Phi g^{-1} \ ,  
$$
where, $g$ may be written as: ${\rm exp}({i\alpha^a \lambda_a})$
for any set of $\alpha^a$.

Now the center\footnote{The center of a group consists of elements 
that commute with all other elements.} of $SU(3)$ is $Z_3$ and the
elements of $Z_3$ are of the form: 
$$
e^{i2\pi n/3} {\bf 1} \ , \ \ n={\rm integer} \ ,
$$ 
where ${\bf 1}$ is the identity matrix. Hence, $\Phi$ is left invariant
under transformations belonging to the center. And so the VEVs
of any number of adjoint scalar fields cannot
break the $Z_3$ center.
Then, one way to achieve (\ref{symbreak}) would be to give VEVs 
to as many adjoint fields as necessary to break the $SU(3)$
maximally, that is, only leaving the center unbroken. Indeed, 
one can check that two
adjoints $\Phi_1$ and $\Phi_2$ are sufficient because if,
\be
\Phi_1 = \lambda_{1} \ , \ \Phi_2 = \lambda_{4} \ 
\label{phivev}
\ee
then the only group elements that commute with both $\Phi_1$
and $\Phi_2$ are the ones proportional to ${\bf 1}$, that is,
the elements of the $Z_3$ center. For completeness, in 
Table \ref{table1} we show the various possible VEVs for
$(\Phi_1 , \Phi_2 )$ and the resulting symmetry breaking pattern.

In the next subsection, we will consider an $SU(3)$ invariant
potential for two adjoint scalar fields. As the full potential has
a lot of parameters, we restrict our attention to a certain region
of parameter space. Within this truncated model, we will find that
it is not possible to construct a potential that will lead to the 
VEVs in (\ref{phivev}) at a unique, global minimum. Then it 
will be necessary to introduce a third adjoint field $\Phi_3$.  In
Table \ref{table2} we show the symmetry breaking patterns for
different directions of $(\Phi_1, \Phi_2 , \Phi_3 )$.

In what follows, we will construct a potential that
will have a global minimum when $(\Phi_1 ,\Phi_2 ,\Phi_3 )$
acquire VEVs in the following direction:
\be
(\lambda_{4}, \lambda_{6}, \lambda_{8} ) \ .
\ee
($SU(3)$ rotations of these VEVs will yield the manifold of global
minima.) As shown in Table II, the residual symmetry group will be
$Z_3$ in this case.

\vskip 1 truecm

\begin{center}
\begin{tabular}{|l|l|l|l|l|l|l|l|l|} 
\tableline
\ & 1 & 2 & 3 & 4 & 5 & 6 & 7 & 8 \\ \hline
1 & U1U1 & U(1) & U(1) & $Z_{3}$ & $Z_{3}$ & $Z_{3}$ & $Z_{3}$ & U1U1 \\ \hline
2 & U(1) & U1U1 & U(1) & $Z_{3}$ & $Z_{3}$ & $Z_{3}$ & $Z_{3}$ & U1U1 \\ \hline
3 & U(1) & U(1) & U1U1 & $U(1)$ & $U(1)$ & $U(1)$ & $U(1)$ & U1U1 \\ \hline
4 & $Z_{3}$ & $Z_{3}$ & $U(1)$ & U1U1 & U(1) & $Z_{3}$ & $Z_{3}$ & U(1) \\ \hline
5 & $Z_{3}$ & $Z_{3}$ & $U(1)$ & U(1) & U1U1 & $Z_{3}$ & $Z_{3}$ & U(1) \\ \hline
6 & $Z_{3}$ & $Z_{3}$ & $U(1)$ & $Z_{3}$ & $Z_{3}$ & U1U1 & U(1) & U(1) \\ \hline
7 & $Z_{3}$ & $Z_{3}$ & $U(1)$ & $Z_{3}$ & $Z_{3}$ & U(1) & U1U1 & U(1) \\ \hline
8 & U1U1 & U1U1 & U1U1 & U(1) & U(1) & U(1) & U(1) & SU2U1 \\ 
\tableline
\end{tabular}
\begin{table}[tbp]
\caption{\label{table1}
$SU(3)$ breaking with two adjoint Higgs fields.
The rows and columns label the direction of the VEVs of each of the two
fields while the table entry gives the residual symmetry.
For convenience of notation we have defined: $U1U1 = U(1)\times U(1)$, and
$SU2U1 = SU(2)\times U(1)$}
\end{table}
\end{center}

\vskip 1 truecm

\begin{center}
\begin{tabular}{|l|l|}
\tableline
U(1) & $(\lambda_{1},\lambda_{2},\lambda_{3})$,\,
$(\lambda_{1},\lambda_{2},\lambda_{8})$,\,
$(\lambda_{1},\lambda_{3},\lambda_{8})$,\\
& $(\lambda_{2},\lambda_{3},\lambda_{8})$,\, 
$(\lambda_{3},\lambda_{4},\lambda_{8})$,\,
$(\lambda_{3},\lambda_{5},\lambda_{8})$,\\
& $(\lambda_{3},\lambda_{6},\lambda_{8})$,\,
$(\lambda_{3},\lambda_{7},\lambda_{8})$,\,
$(\lambda_{4},\lambda_{5},\lambda_{8})$,\\
& $(\lambda_{6},\lambda_{7},\lambda_{8})$,\, \\ \hline
$Z_{3}$ & all other  $(\lambda_{i},\lambda_{j},\lambda_{k})$
with distinct $i,j,k$ \\
\tableline
\end{tabular}
\begin{table}[tbp]
\caption{\label{table2} $SU(3)$ symmetry breaking with three distinct
adjoint scalar fields. The first column shows the residual symmetry group
if the fields get VEVs in the directions shown in the second column.}
\end{table}
\end{center}

\subsection{Construction of the potential}

The $SU(3)$ invariant potential for three adjoint
fields can be written as follows:
\be
V (\{\Phi_l\}) = V_1 (\{\Phi_l\}) + V_2 (\{\Phi_l\}) \ ,
\label{fullpot}
\ee
where
\be
V_1 (\{\Phi_l\}) = \sum_{l=1}^3 [ - m_l^2 ({\rm Tr} \Phi_l^2)
             +a_l ({\rm Tr} \Phi_l^2)^2 +b_l {\rm Tr}\Phi_l^4 ]
\ee
and
\begin{eqnarray}
V_2 (\{\Phi_l\}) = \sum_{l=1}^3 [ c_l {\rm Tr} (\Phi_m \Phi_n)
                        + d_l ({\rm Tr}(\Phi_m \Phi_n ))^2 \\ \nonumber
             +e_l {\rm Tr}(\Phi_m \Phi_n)^2
             +f_l {\rm Tr}(\Phi_m^2 \Phi_n^2) \\ \nonumber
              +g_l {\rm Tr}(\Phi_l^2\Phi_m \Phi_n )
              +h_l {\rm Tr}(\Phi_m \Phi_l^2 \Phi_n ) ]
\end{eqnarray}
with, $l,m,n$ taking cyclic values over $1,2,3$. (In writing the
potential we have omitted cubic terms for simplicity.)

The full Lagrangian can now be written: 
\be
L = \sum_l  {\rm Tr}(| D_\mu \Phi_l | ^2 )
  - {1\over 2} {\rm Tr}(G_{\mu \nu} G^{\mu \nu}) - V(\{ \Phi_l \})
\label{model}
\ee
where,
$$
D_\mu \Phi_l \equiv \partial_\mu \Phi_l + g [A_\mu , \Phi_l ] \ ,
$$
$A_\mu$ is the matrix-valued gauge field,
$$
G_{\mu \nu} = G_{\mu \nu}^a \lambda_a
            = \partial_\mu A_\nu - \partial_\nu A_\mu
               +g [A_\mu , A_\nu ] \ .
$$

The general potential (\ref{fullpot}) has 27 parameters and is far
too complicated for us to handle. We will assume
certain relationships among the parameters to enable us
to proceed further. These are:
$$
a_l =a ~ , ~  b_l = b ~ , ~ m_l = {m \over 2} ~ ,
$$
and
$$
c_l = 0 =d_l = g_l = h_l ~ .
$$

Now to check if two scalar fields would have been sufficient for
the symmetry breaking (\ref{symbreak}), we eliminate all terms
containing $\Phi_3$ in (\ref{fullpot}), restrict our attention to
the truncated region in parameter space, then feed in the various
possible VEVs for $\Phi_1$ and $\Phi_2$ from Table \ref{table1}.
We find that the VEVs leading to a $Z_3$ residual symmetry 
give a higher (or equal) energy than the other symmetry breaking 
patterns. So the $Z_3$ vacuum cannot be a unique global minimum. Hence, 
it is necessary for us to include the third scalar field $\Phi_3$.

With three scalar fields, we introduce 
$$
\Lambda = 16 a + 8 b ~ ,
$$
and, make the further choice of parameters:
$$
\epsilon = f_l = e_1 = e_2 = -e_3 ~ 
$$
where, $\epsilon$ is a new free parameter.
The minus sign in front of $e_3$ is a crucial feature that
ensures that the global minimum of the potential has $Z_3$
symmetry.

The potential for this restricted set of parameters is
\begin{eqnarray}
{\bar V} (\{\Phi_l\}) = 
          \sum_{l=1}^3 &[& - {{m^2}\over 4} ({\rm Tr} \Phi_l^2) 
             +a ({\rm Tr} \Phi_l^2)^2 +b {\rm Tr}\Phi_l^4 ] \nonumber \\
  +\epsilon &[& {\rm Tr}(\Phi_2 \Phi_3)^2
            + {\rm Tr}(\Phi_3 \Phi_1)^2   
            - {\rm Tr}(\Phi_1 \Phi_2 )^2 \nonumber \\
            &+& {\rm Tr}(\Phi_2^2 \Phi_3^2) 
            + {\rm Tr}(\Phi_3^2 \Phi_1^2) 
            + {\rm Tr}(\Phi_1^2 \Phi_2^2) ] ~ . 
\label{vbar}
\end{eqnarray}
For the potential to have a global minimum at finite VEVs of the
fields, we need
$$
\Lambda > 0 ~ .
$$
The requirement that the VEVs of the fields be non-vanishing 
gives the constraint:
$$
{\epsilon \over \Lambda} < {3\over 2} ~ .
$$ 
Within this parameter range, we have inserted
all possible choices of directions of $\Phi_l$ ($l=1,2,3$)
in the potential and find that it has a 
global minimum when the residual symmetry group is $Z_3$, provided
$$
\epsilon > 0 ~ .
$$
Further, the VEVs yielding the global minimum are in the 
$(\lambda_4 , \lambda_6 , \lambda_8 )$ directions (and $SU(3)$
rotations of these directions):
\begin{eqnarray}
\Phi_1 &=& \eta_1 \lambda_{4} \nonumber \\ 
\Phi_2 &=& \eta_2 \lambda_{6} \\
\label{phivevs}
\Phi_3 &=& \eta_3 \lambda_{8} \nonumber
\end{eqnarray}
where
\begin{equation}
\eta_1=\eta_2 = m \sqrt{ {{\Lambda -2\epsilon /3} \over
                     {\Lambda^2 +2\epsilon \Lambda -8\epsilon^2/9}}}
\label{phi12vev}
\end{equation}
\begin{equation}
\eta_3 =  m \sqrt{ {{\Lambda +2\epsilon /3} \over
                     {\Lambda^2 +2\epsilon \Lambda -8\epsilon^2/9}}} ~ .
\label{phi3vev}
\end{equation}

Let us now define:
$$
v_l^2 \equiv {1 \over 2}{\rm Tr}( \Phi_l^2 ) ~ . 
$$
Then, for a vortex, $v_l$ will vary in space and the relevant
potential is:
\begin{eqnarray}
V(\{v_l\}) = \sum_l \biggr [ &-& {{m^2}\over 2} v_l^2
           + {\Lambda \over 4} v_l^4 \biggr ] \nonumber \\
             &+& {\epsilon \over 3} (v_1^2 + v_2^2) v_3^2
               +\epsilon v_1^2 v_2^2 ~ .
\label{reducedpot}
\end{eqnarray} 
\section{$Z_3$ String Ansatz and Solution}
\label{solutions}

Having truncated the full model to one which is simple
enough to analyze, we now write down the ansatz for
$Z_3$ strings, insert it into
the field equations and then find
the string solutions.

To write a string ansatz, we must first specify a closed path
on the vacuum manifold, $P (\theta )$, parametrized by 
$\theta \in [0,2\pi ]$, which is incontractable. This is given by:
$$
P(\theta ) = e^{i n \lambda_8 \theta /{\sqrt{3}}} \ .
$$
This path is incontractable since $P(2\pi )$ is a non-trivial element
of the discrete residual group $Z_3$ for $n=1,2$.

We now identify $\theta$ with the spatial polar coordinate. 
Then the scalar field ansatz is:
$$ 
\Phi_l (r \rightarrow \infty , \theta ) = 
P(\theta )^{\dag} \Phi_l (r \rightarrow \infty , \theta =0 ) P(\theta ) \ .
$$
With $\Phi_l (\theta =0)$ given by eq. (\ref{phivevs}),
this leads to:
\begin{eqnarray}
\Phi_{1} (r,\theta ) &=& v_{1}(r) (cos\,n\theta ~\lambda_{4}
        + sin\,n\theta ~\lambda_{5}) \\ \nonumber
\Phi_{2} (r,\theta ) &=& v_{2}(r) (cos\,n\theta ~\lambda_{6}
        + sin\,n\theta ~\lambda_{7}) \\ \nonumber
\Phi_{3} (r,\theta ) &=& v_{3}(r) \lambda_{8} \\ \nonumber
A_{\theta}^{8} &=& -{n \over {\sqrt{3} g}}{\alpha(r) \over r}
\end{eqnarray}
where we have included scalar field profile functions $v_l (r)$
and given the gauge field ansatz with its profile function
$\alpha (r)$. All other components of the gauge field are taken
to vanish.

We now insert this ansatz into the field equations to get:
\begin{eqnarray*}
v_{1}''+{1\over r} v_{1}'-{n^2\over r^2}(1-\alpha)^2 v_{1}  
-\Lambda v_{1}^3 + m^2 v_{1} \\
-2\epsilon(v_{2}^2+ {1\over 3} v_{3}^2) v_{1} &=& 0        \\ 
v_{2}''+{1\over r} v_{2}'-{n^2\over r^2}(1-\alpha)^2 v_{2}  
-\Lambda v_{2}^3 + m^2 v_{2} \\
-2\epsilon(v_{1}^2+ {1\over 3} v_{3}^2) v_{2} &=& 0       \\ 
v_{3}''+{1\over r} v_{3}' -\Lambda v_{3}^3 + m^2 v_{3}
-{2\over 3}\epsilon (v_{1}^2+v_{2}^2) v_{3} &=& 0           \\ 
\alpha''-{1\over r}\alpha'+3g^{2}(v_{1}^{2}+v_{2}^2)(1-\alpha)
&=& 0  
\end{eqnarray*}
where a prime denotes differentiation with respect to $r$.

The equations for $v_1$ and $v_2$ are identical and so are the 
asymptotic boundary conditions as is seen from (\ref{phi12vev}).
Therefore we set
$$
v_1 (r) = v_2 (r) \ .
$$

It is now convenient to define rescaled coordinates, parameters and 
fields as follows:
\begin{equation}
x = g \sqrt{6 \over \Lambda} m r
\label{xandr}
\end{equation}
$$
\lambda^2 = {\Lambda\over {3 g^2}} \ , \ \ \
\beta = {\epsilon \over {3g^2}}
$$
$$
v_{1}=v_{2}={m \over \sqrt{\Lambda}} f(x) \ , \ \ \ 
v_{3}={m \over \sqrt{\Lambda}} h(x)
$$
Primes will now denote differentiation with respect to the 
rescaled coordinate $x$.

The rescaled equations are:
\begin{eqnarray}
f'' +{{f'}\over x}-{n^2 \over x^2}(1-\alpha)^2 f -  
{\lambda^2\over 2}(f^2-1)f - \nonumber \\
\beta (f^2+{{h^2}\over 3})f =0 
\label{eom1}
\end{eqnarray}
\begin{equation}
h''+{{h'}\over x}-{\lambda^2\over 2}(h^2-1)h - 
{{2\beta}\over 3}f^2 h = 0 
\label{eom2}
\end{equation}
\begin{equation}
\alpha''-{{\alpha '}\over x}+ f^2 (1-\alpha)=0 ~ .
\label{eom3}
\end{equation}

The boundary conditions on the functions $f$, $h$ and $\alpha$
follow by requiring single-valuedness and regularity of the
fields at the origin, 
\begin{equation}
\alpha(0)=f(0)=h' (0)=0 \ .
\label{afh0}
\end{equation}
At infinity, the fields should go to their vacuum expectation
values:
\begin{equation}
\alpha({x\to \infty})=1 ~ , ~ 
f({x\to \infty}) = F_0 ~ , ~ h({x\to \infty})=H_0  
\label{afhinfty}
\end{equation}
with,
\begin{equation}
F_{0}=\sqrt{{1-2\sigma /3}\over {1+2\sigma- 8\sigma^2 /9}} ~ , ~ 
H_{0}=\sqrt{{1+2\sigma /3}\over {1+2\sigma- 8\sigma^2 /9}}   ~ ,
\label{fhasymptotic}
\end{equation}
where 
$$
\sigma \equiv {\beta \over {\lambda^2}} ~ .
$$
If we set $\beta =0$, the $h$ equation is solved by $h=1$
and the $f$ and $\alpha$ equations are exactly the Nielson-Olesen 
equations for the Abelian-Higgs vortex. These vortices have been
studied extensively and this is also the case discussed by de Vega
and Schaposhnik \cite{dvs}. The interaction of Abelian-Higgs vortices
is characterized by the single ratio of length scales entering
the problem - namely, the ratio of the (single) scalar and vector masses.
With $\beta \ne 0$, however, the picture is more complicated since we have 
three length scales corresponding to the masses of the two independent 
scalar fields in our truncated model, and the gauge fields.
Therefore there are two independent dimensionless 
ratios we can construct, and a correspondingly richer 
structure to the interaction of $Z_3$ vortices\footnote{In the
full model with 27 parameters, one might expect the interaction 
to be even more complex. However, our understanding of the results
presented in Sec. \ref{discuss} indicate that the picture is likely 
to be quite simple even in the full model.}.

\begin{figure}[tbp]
\caption{\label{profile}
The profile functions $f$, $h$ and $\alpha$ for the unit winding
$Z_3$ vortex for $\lambda = 1.13$ and $\beta = 0.42$.
}

\

\epsfxsize = \hsize \epsfbox{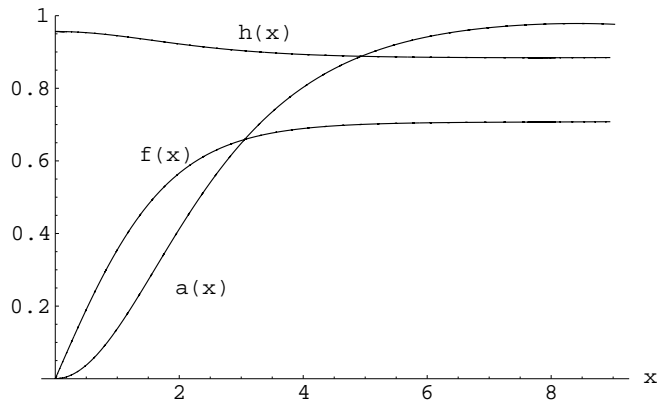}
\end{figure}

We have solved the equations of motion (\ref{eom1})-(\ref{eom3}) 
by a numerical shooting routine.
In Fig. \ref{profile} we show the behaviour of the $f$, $h$ and $v$ fields
for a particular choice of parameters and for $n=1$. 

The equations of motion for the $n=2$ vortex can also
be solved. So the $n=2$ vortex is indeed a solution, though it will be 
unstable to decay into the topologically equivalent $n=-1$ solution
which has lower energy. This instability of the $n=2$ vortex is
not relevant for us since all that we are interested in is the energy
of two overlapping $n=1$ vortices which is the same as the energy of
the $n=2$ vortex solution.

\section{Energy Functional and Numerical Evaluation}
\label{interact}

The set of field equations (\ref{eom1})-(\ref{eom3}) 
can also be obtained by extremizing the energy 
functional 
\be
E= \int d^{3}x~[{1\over 4}G_{ij}^{a}G^{aij} + {1\over 2}
	(D_{i}\Phi^a_{l})(D_{i}\Phi^a_{l}) + V(\{\Phi_{l}\})]
\ee
(The index $a=1,...,8$ is the group index and $l=1,2,3$ labels
the different adjoint fields.) Here we will obtain the extremum
values of the energy for the $n=1,2$ topological configurations
directly, that is, without solving the field equations of motion.
This is the technique used by Jacobs and Rebbi \cite{ljcr} 
to study the interaction of $U(1)$ vortices, and we will employ
it to study the interaction of $Z_3$ vortices. 

Let us first define
\begin{equation}
\mu=\lambda\sqrt{1-{2\over 3}\sigma} ~ , ~ 
\nu=\lambda \sqrt{{1+2\sigma /3}\over {1+2\sigma- 8\sigma^2 /9}} ~ .
\label{muandnu}
\end{equation}
Then the solution to eq. (\ref{eom1})-(\ref{eom3}) may be written 
as:
\be
f=F_{0}\biggr ( 1+\sum_{j=0}^{\infty}{f_{j}\over j!}x^{j} e^{-\mu x}
       \biggr ) 
\ee
\be
h=H_{0}\biggr (1+\sum_{j=0}^{\infty}{h_{j}\over j!}x^{j} e^{-\nu x}
       \biggr ) 
\ee
\be
\alpha=1+\sum_{j=0}^{\infty}{\alpha_{j}\over j!}x^{j} e^{-x} \ .
\ee
In this form, the functions automatically satisfy the desired boundary
at infinity (eq. (\ref{afhinfty})). The boundary conditions at the origin
(eq. (\ref{afh0})) require
$$
\alpha_{0}=-1~ , ~ f_{0}=-1 ~ , ~ h_1=\nu h_0 ~ , 
$$
for both $n=1$ and $n=2$ vortices. In addition, regularity of the
gauge fields at the origin requires
$$
\alpha ' (0) =0
$$
and,
$$
f '(0) =0 \, , \ {\rm for} ~ n=2 ~ .
$$
These conditions give
$$
\alpha_1 = -1
$$
and 
$$
f_1 = - \mu \, , \ {\rm for} ~ n=2 ~ .
$$

In this scheme, since we are numerically evaluating the energy functional,
it is necessary to ensure that the potential
vanishes in the true vacuum. This requires that we shift the potential
in (\ref{reducedpot}) by a constant $v_0$:
\begin{eqnarray*}
v_{0}=-(2 F_{0}^4+H_{0}^4)+2(2F_{0}^2+H_{0}^2) - 
        {8\over 3}\sigma F_{0}^2 H_{0}^2 +4 \sigma F_{0}^4 \ .
\end{eqnarray*}
The potential can now be written in terms of the fields $f$ and $h$:
\begin{eqnarray*}
V(f,h)={\lambda^2 \over 8}\biggr [(2f^4+h^4)-2(2f^2+h^2) \\
            +{8\over 3}\sigma f^2 h^2 + 4\sigma f^4 + v_0 \biggr ]
\end{eqnarray*}

In terms of the fields $f$, $h$ and $\alpha$, the energy
is:
$$
E = {{2 \pi m}\over {\sqrt{6\Lambda} g}} 
\int dz ~ dx ~ x ~ {\cal E} [f,h,\alpha ]
$$
where $z$ is the rescaled (as in eq. (\ref{xandr})) 
dimensionless coordinate along the string, and
\begin{eqnarray*}
{\cal E} [f,h,\alpha ] = f'^2 + {1\over 2} h'^2 
                        + n^2 \left ({\alpha' \over x} \right )^2
       &+& {n^{2}\over {x^2}}(1-\alpha)^2 f^2   \\
       &+& V(f,h) ~ .
\end{eqnarray*}

We now have to evaluate the energy functional in terms of the various 
coefficients $f_i$, $h_i$ and $\alpha_i$ (infinite in number), for 
different choices of $\{\lambda,\beta \}$ and winding number $n$. 
On inserting the expansions for the fields in the 
energy functional, we end up with integrals that can be evaluated
in terms of Gamma functions. These form the coefficients of the
quartic order polynomials in the $f_i$, $h_i$ and $\alpha_i$.
This gives $E = E (\{f_{i}\}, \{h_{i}\}, \{\alpha_{i}\})$ where
$i=0,...,\infty$. 
We then truncate the expansion by retaining only 
the first 7 terms in each of the expansions ($i=0,...,6$), making a 
total of 21 parameters that need to be varied to minimize $E$.
We finally find the global minimum of $E$ with
respect to the variation of the coefficients for $n=1,2$ and
$\lambda \in (0,10)$ and $\beta \in (0,6)$.

\begin{figure}[tbp]
\caption{\label{evsbeta}
The dependence of the energy of $n=1$ and $n=2$ vortices
on the parameter $\beta$ for $\lambda = 2.0$. Also shown
is twice the energy of the $n=1$ vortex. 
}

\

\epsfxsize = \hsize \epsfbox{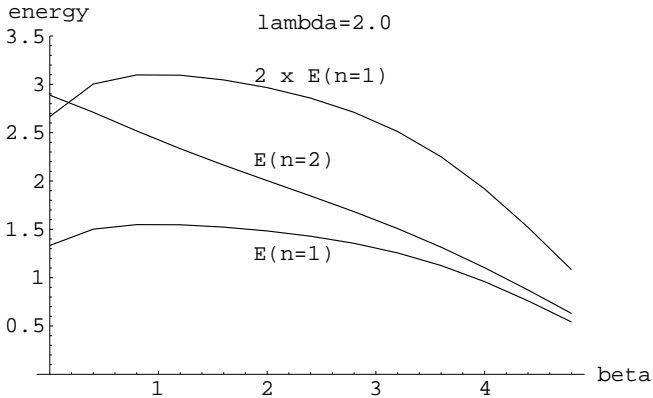}
\end{figure}

Sample results for the dependence of the energy on the $\beta$ parameter 
are shown in Fig. \ref{evsbeta}.

\section{Critical Couplings and Inter-Vortex Forces}
\label{critical}

In the $\beta =0$ limit, the equation of motion (\ref{eom2})
is simply solved by 
$$
h(x; \beta =0) = 1
$$
and the remaining equations (\ref{eom1}) and (\ref{eom3})
are identical to the $U(1)$ equations. So, in this case,
the structure and interaction of $Z_3$ vortices 
is identical to those of $U(1)$ vortices.
In particular, for $\lambda =1$, 
the inter-vortex potential vanishes \cite{dvs}. If, however,
we now return to the potential (\ref{vbar}) and set
$\epsilon =0$ (equivalent to $\beta =0$), we find that
the potential does not uniquely pick out the required directions
for the symmetry breaking (\ref{symbreak}). Indeed, the VEVs
of the three different fields can point in the same direction,
say in the $\lambda_8$ direction, leading to an
$SU(2)\times U(1)$ residual symmetry. So to pick out the desired
symmetry breaking we must necessarily consider $\beta \ne 0$.

\begin{figure}[tbp]
\caption{\label{criticalcurve}
The curve in parameter space for which the energy of two inifintely
separated vortices have the same energy as two vortices at zero
separation, that is, the $n=2$ vortex. The region in which the
$n=2$ vortex is more energetic than two $n=1$ vortices is the 
``repulsive'' region while that in which the $n=2$ vortex is less 
energetic is the ``attractive'' region.
}

\

\epsfxsize = \hsize \epsfbox{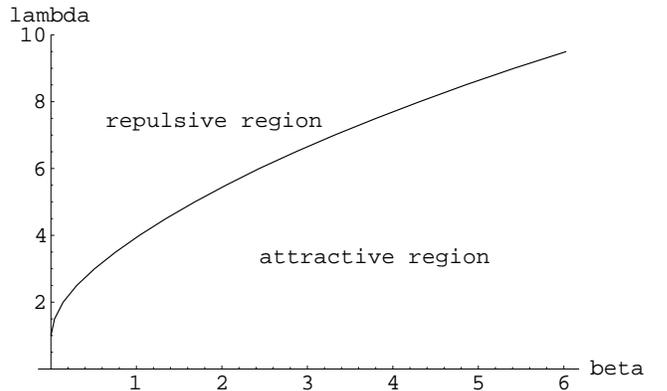}
\end{figure}

As one can see in Fig. \ref{evsbeta}, there are points in parameter
space where
\begin{equation}
E(\beta , \lambda ; n=2) = 2 \times E(\beta , \lambda ; n=1) \ .
\label{2e1eqe2}
\end{equation}
In this case, the energy of two infinitely separated vortices
is equal to the energy of two overlapping vortices. We shall
call the parameters for which (\ref{2e1eqe2}) holds, to be
``critical''. In Fig. \ref{criticalcurve} we plot the critical curve in 
$(\beta , \lambda )$ space.

The question we now address is whether the critically coupled vortices
have a mutual repulsion or attraction at intermediate separations.
In the critically coupled $U(1)$ case ($\beta =0$, $\lambda =1$), 
the intervortex potential is flat and hence the intervortex forces
vanish at all separations. It is useful to think of this in terms
of the repulsive gauge field and attractive scalar field interactions.
The exchange of spin one gauge particles between identical vortices
leads to a repulsive force while the exchange of spin zero particles
leads to an attractive force. So the gauge field repulsion is balanced
by the scalar field attraction in the critical $U(1)$ case.

\begin{figure}[tbp]
\caption{\label{intervortexpotential}
A schematic depiction of the expected intervortex potential ($U$) 
as a function of vortex separation ($s$)
in the critically coupled case where the gauge field mass is
less than the scalar field masses. Being lighter, the gauge field
provides a longer range interaction than the scalar fields and
leads to a repulsive force between vortices in the asymptotic
region. In a $U(1)$ model, 
the single scalar field is unable to overcome the repulsive potential 
at short distances. However, in the $Z_3$ string case, the two 
scalar fields successfully turn the potential over at short distances 
leading to a maximum at $s=s_*$.
}

\

\epsfxsize = \hsize \epsfbox{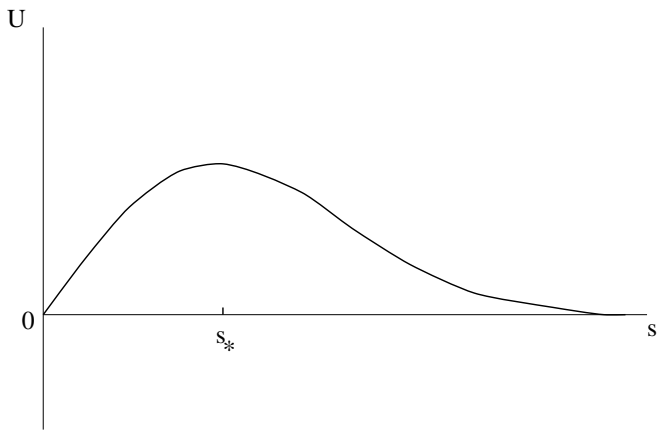}
\end{figure}

In the case of the $Z_3$ string, the crucial observation is that
the curve of critical couplings lies in the region $\lambda \ge 1$
and small $\sigma = \beta /\lambda^2$.
Hence, along the critical curve,
the gauge field mass is smaller than either of the scalar
field masses $\mu$ and $\nu$ given in (\ref{muandnu}).
So the gauge field interaction is of longer range than the
scalar field interaction. Therefore, if we bring in two infinitely
separated vortices, they will first experience the repulsive gauge
field interaction. When they come in closer, the attractive 
interaction due to the scalar fields will turn on. Since there are
two scalar fields in our model, the attraction is stronger than in
the $U(1)$ case and is more effective in cancelling out the gauge
field repulsion. So the intervortex potential is expected to turn 
over, as schematically depicted in Fig. \ref{intervortexpotential}. 
For the critically coupled
case, the turn over is such that the energy at zero vortex separation 
equals that at infinite vortex separation. The presence of a turning
point in the intervortex potential at some vortex separation $s_*$ 
means that two vortices separated by this distance can be in 
relative equilibrium. In the present case, this is an unstable
equilibrium because the potential is a maximum.

Note that this argument is based on the fact that the curve of
critical couplings lies in the $\lambda \ge 1$ region. This feature
can be understood by realizing that the presence of two scalar fields 
instead of one implies an enhanced attractive force between vortices
at short distances. So if one imagines starting out with repulsive 
$U(1)$ vortices, and adding a second scalar field as is present in
the $Z_3$ vortices, this can turn over
the intervortex potential at small separations and make the energy
at $s=0$ the same as that at $s=\infty$. But since one needs to start
out with repulsive $U(1)$ vortices, this means that critically coupled
$Z_3$ vortices should lie in the $\lambda \ge 1$ region. (If one started
out with attractive ($\lambda < 1$) $U(1)$ vortices, the addition of a 
second scalar field would simply make the vortices even more attractive 
at short distances and eq. (\ref{2e1eqe2}) could never be satisfied.)

\section{Conclusions}
\label{discuss}

We have constructed field theory solutions for $Z_3$
strings and have studied their interaction within a
range of model parameters. The solution for the structure
of the vortex is shown in Fig. \ref{profile} while the dependence 
of the energy on the new parameter in the model, that is, the
parameter not present in the $U(1)$ case, is shown in Fig. \ref{evsbeta}.
Infinitely separated $Z_3$ strings have the same energy as 
strings at zero separation along a curve in our two dimensional space 
of parameters (Fig. \ref{criticalcurve}).
However, we have given general arguments to show that,
unlike the $U(1)$ case, the intervortex potential is not trivial 
for these critically coupled $Z_3$ strings. In fact, the intervortex
potential is expected to have a maximum value at some
non-vanishing vortex separation (Fig. \ref{intervortexpotential}). 
This suggests that an 
{\it unstable} bound state of two $Z_3$ vortices should exist.

Although we have worked in detail within a specific range of
parameters, we have understood the intervortex forces based on 
the number of scalar and vector fields present in the model and 
their masses. This reasoning (described in Sec. \ref{critical})
is expected to be valid quite generally 
and should apply to the full range of parameters in this model as 
well as to other models.

\

\

\noindent{\it Acknowledgements:} We are grateful to Matt Strassler 
for discussions regarding ongoing work on supersymmetric dualities
suggesting that $Z_3$ strings may be relevant to confinement in QCD,
and also for bringing Ref. \cite{dvs} to our attention.
We would also like to thank Dimitri Kusnezov, Patrick McGraw
and Grisha Volovik for helpful comments.
TV is supported by a research grant from the DoE.

\end{document}